\documentclass[reprint, superscriptaddress, amsmath, amssymb, aps, prb]{revtex4-2}
\usepackage[dvipdfmx]{graphicx}
\usepackage{dcolumn}
\usepackage{bm}
\usepackage[dvipdfmx]{hyperref}
\usepackage{comment}
\usepackage{color}
\usepackage{physics}
\begin{document}
\preprint{APS/123-QED}
\title{Spin-polarized Specular Andreev Reflections in Altermagnets}
\author{Yutaro Nagae}
\thanks{These authors contributed equally to this work.}
\affiliation{Department of Applied Physics, Nagoya University, Nagoya 464-8603, Japan}
\author{Andreas P. Schnyder}
\affiliation{Max-Planck-Institut f\"ur Festk\"orperforschung, Heisenbergstrasse 1, D-70569 Stuttgart, Germany}
\author{Satoshi Ikegaya}
\thanks{These authors contributed equally to this work.}
\affiliation{Department of Applied Physics, Nagoya University, Nagoya 464-8603, Japan}
\affiliation{Institute for Advanced Research, Nagoya University, Nagoya 464-8601, Japan}
\date{\today}

\begin{abstract}
We show theoretically that specular Andreev reflection occurs stably at altermagnet--superconductor interfaces,
which is a phenomenon that has previously been predicted only in a limited range of materials, such as Dirac/Weyl materials with fine-tuned chemical potentials.
Furthermore, the characteristic spin-split bands of the altermagnet lead to a distinctive spin polarization in the specular Andreev reflections.
By utilizing this feature, we propose a device that integrates the functions of both a Cooper pair splitter and a spin beam splitter, thereby creating energy-entangled electron pairs.
The positive nonlocal conductance and the positive noise cross-correlation are unambiguous signatures of specular Andreev reflections in the proposed device.
\end{abstract}
\maketitle

\section{Introduction}.~%
Andreev reflection (AR) is a fundamental scattering process that occurs at normal metal--superconductor (NM--SC) interfaces~\cite{andreev_64};
in this process an electron injected from the NM is reflected as a hole, while a Cooper pair is transmitted into the SC.
Usually, ARs exhibit a retroreflective nature, i.e., the reflected holes retrace the path of the incident electrons.
In contrast, there are two representative \emph{non-retroreflective} ARs: crossed AR (CAR)~\cite{deutscher_00,loss_01} and specular AR (SAR)~\cite{beenakker_06}.
In CAR, an incident electron from one lead is transmitted as a hole in another lead spatially separated by the SC.
In SAR, an incident electron is \emph{specularly} reflected as a hole due to the particular band structure of the material forming the NM segment.
Remarkably, the reversal process of these non-retroreflective ARs, in which Cooper pairs are injected from the SC to the NM, corresponds to Cooper pair splitting.
The Cooper pair splitting has attracted considerable interest in the broader field of quantum mechanics,
as it opens a promising avenue for the creation of non-local entanglement in \emph{electron} pairs~\cite{martin_99,blatter_01,martin_02,buttiker_03},
an essential element for future quantum devices enabling computing, sensing and data processing in the quantum domain~\cite{loss_98,vidal_03}.

In this Letter, we theoretically address two significant issues related to the non-retroreflective ARs.
(i) First, we address the challenge for definitive observations of SAR.
To date, SAR has only been predicted for a narrow range of materials, specifically Dirac/Weyl materials with the relativistic band dispersion~%
\cite{beenakker_07,silvestrov_07,asano_08,xing_08,sun_09,ma_12,recher_12,chen_13,asgari_16,sun_17,danneau_19,sun_20,sun_21,belzig_23}.
Furthermore, SAR in Dirac/Weyl materials occurs predominantly only in an extreme regime
where the chemical potential in the Dirac/Weyl material is much smaller than the pair potential in the SC~\cite{beenakker_06}.
Due to this difficulty, experimental studies of SAR have been limited~\cite{komatsu_12,kim_16,das_16}, and definitive observations of this phenomenon have not yet been achieved.
(ii) Second, we address the challenge of creating a specific entanglement in electrons.
In the seminal paper proposing the creation of entangled electrons through Cooper pair splitting, two distinct types of entanglement are discussed:
spin (energy) entanglement created by Cooper pair splitter equipped with energy (spin) filters~\cite{blatter_01}.
Nevertheless, the previous studies have focused exclusively on the spin entanglement induced by CARs in hybrid devices consisting of a SC and quantum dots acting as energy filters~%
\cite{schonenberger_09,chandrasekhar_09,chandrasekhar_10,schonenberger_11,schonenberger_12,shtrikman_12,hakonen_15,csonka_15,tarcha_16}.
In contrast, the physics of energy-entangled electrons remains largely unexplored due to the lack of suitable systems to observe Cooper pair splitting in the presence of spin filters.
Here, we propose a promising device addressing both of these significant issues, i.e., (i) and (ii).

\begin{figure}[bbbb]
\begin{center}
\includegraphics[width=0.38\textwidth]{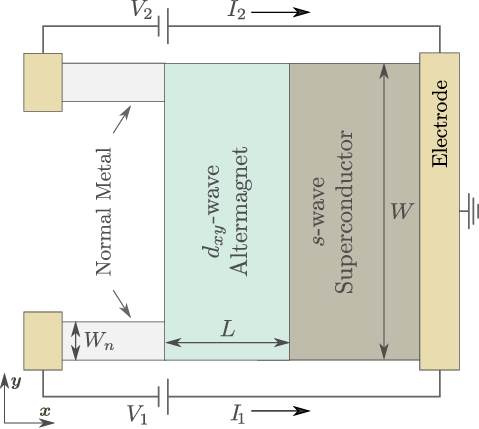}
\caption{Schematic image of the device consisting of an AM, a $s$-wave SC, and two NM leads.}
\label{fig:figure1}
\end{center}
\end{figure}
A key element of our proposal is the use of an altermagnet (AM), a recently discovered magnetic material characterized by the coexistence of spin-split bands and compensated collinear magnetic order~%
\cite{seo_19,hayami_19,hayami_20,smejkal_20,tsymbal_21,smejkal_22(0),smejkal_22(1),smejkal_22(2),mazin_22,yao_24,elmers_24,smejkal_24,%
zunger_20,spaldin_24,mazin_21,reichlova_24,leivisk_24,tang_23,liu_24,qian_24}.
Currently, the physics of AM--SC hybrid devices is a vibrant area of research,
as demonstrated by the numerous works that have appeared just within the last few months,
discussing ARs~\cite{linder_23,papaj_23,soori_24,wang_24}, proximity effects~\cite{neupert_23,linder_23(2),zyuzin_24,belzig_24},
Josephson effects~\cite{linder_23(3),beenakker_23,sun_23,tanaka_24}, superconducting diode effects~\cite{scheurer_24,sum_24},
superconducting spin-splitting effects~\cite{linder_24}, and topological superconductivity~\cite{yan_23,liu_23,sudbo_23,cano_23}.
In this Letter, we show that SARs occur at interfaces between a $s$-wave SC and a $d$-wave AM,
where the $d$-wave altermagnetism is expected to emerge in various candidate materials, including
$\kappa$-Cl~\cite{seo_19}, RuO$_2$~\cite{smejkal_20,concern_ruo2}, MnF$_2$~\cite{zunger_20,spaldin_24},
FeSb$_2$~\cite{mazin_21}, Mn$_5$Si$_3$~\cite{reichlova_24,leivisk_24},
twisted antiferromagnetic bilayer systems~\cite{tang_23,liu_24}, and KV$_2$Se$_2$O~\cite{qian_24}.
In particular, we show that the SARs in the AM exhibit a distinctive spin polarization that cannot be observed in Dirac/Weyl materials.
Taking advantage of these unique properties of spin-polarized SARs, we propose a multiterminal device that integrates the functions of both a Cooper pair splitter and a spin beam splitter.
Importantly, our proposed AM-based Cooper pair splitter inherently incorporates spin filters, making it a promising system for creating energy-entangled electron pairs.
Experiments to directly prove the energy entanglement (such as Bell's inequality violation experiments) have not yet been established.
Nevertheless, positive nonlocal conductance and positive noise cross-correlation are unambiguous fingerprints of SAR in the proposed device.

\section{Model and Formulation}.~%
\begin{figure}[t]
\begin{center}
\includegraphics[width=0.45\textwidth]{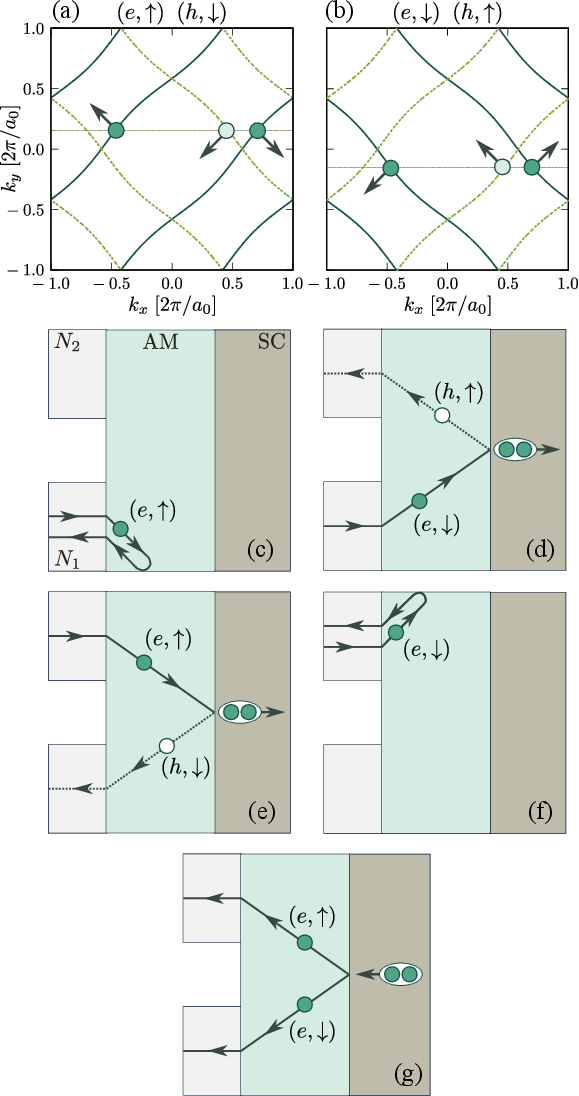}
\caption{(a) [(b)] FSs for the spin-$\uparrow$ (spin-$\downarrow$) electron (solid line) and the spin-$\downarrow$ (spin-$\uparrow$) hole (dashed line) in the AM.
The FSs are obtained at $t_J=0.75t_0$ and $\mu_0=0.5t_0$.
(c) and (d) [(e) and (f)] Scattering processes in which spin-$\uparrow$ and spin-$\downarrow$ electrons are injected from the first (second) NM, respectively.
(g) Spin-polarized Cooper pair splitting corresponding to the reversal processes of the spin-polarized SARs in (d) and (e).}
\label{fig:figure2}
\end{center}
\end{figure}
We consider a hybrid device shown in Fig.~\ref{fig:figure1}, which consists of an AM, a $s$-wave SC, and two NM leads, on a two-dimensional tight-binding model.
For the AM segment, we assume a metallic AM with a $d_{xy}$-wave magnetic order with spin-split open Fermi surfaces (FSs),
which occur, for example, in the $d$-wave altermagnets RuO$_2$ and KV$_2$Se$_2$O~\cite{elmers_24,qian_24}.
Specifically, as shown in Fig.~\ref{fig:figure2}(a) [Fig.~\ref{fig:figure2}(b)], the spin-$\uparrow$ (spin-$\downarrow$) FS has an open shape along the $(k_x+k_y)$ [$(k_x-k_y)$] direction.
Here, the $x$ and $y$ directions of our model correspond to the $[100]$ and $[010]$ ($[110]$ and $[\bar{1}10]$) crystal orientations of RuO$_2$ (KV$_2$Se$_2$O), respectively.
In order to reproduce such spin-split open FSs within a versatile minimal model,
we employ an effective single-band model with a $d_{xy}$-wave exchange potential, as used in previous studies~\cite{smejkal_22(1),linder_23(2),linder_23(3)},
\begin{align}
\begin{split}
&H_{\mathrm{AM}}=\sum_{\boldsymbol{k},\sigma} \left( \varepsilon_{\boldsymbol{k}} - \mu_0
+ s_{\sigma} m_{\boldsymbol{k}} \right) c^{\dagger}_{\boldsymbol{k}\sigma}c_{\boldsymbol{k}\sigma},\\
&\varepsilon_{\boldsymbol{k}}=-2t_0 \cos \frac{k_x}{2} \cos\frac{k_y}{2}, \\
&m_{\boldsymbol{k}} = -2t_J \sin\frac{k_x}{2} \sin\frac{k_y}{2},
\label{eq:ham_am_k}
\end{split}
\end{align}
where $c_{\boldsymbol{k}\sigma}^{\dagger}$ ($c_{\boldsymbol{k}\sigma}$) is the creation (annihilation) operator of an electron
with momentum $\boldsymbol{k}$ and spin $\sigma$ ($\uparrow$ or $\downarrow$),
$t_0>0$ denotes the nearest-neighbor hopping integral on a square-lattice oriented at a $45^\circ$ degree angle,
$\mu_0$ is the chemical potential, $t_J>0$ is the strength of the $d_{xy}$-wave exchange potential, and $s_{\sigma} = +1 (-1)$ for $\sigma=\uparrow$ ($\downarrow$).
For $2t_J>|\mu_0|$ with $t_J<t_0$ and $2t_0>|\mu_0|$ with $t_J>t_0$,
the effective model in Eq.~(\ref{eq:ham_am_k}) exhibits the open FSs as shown in Figs.~\ref{fig:figure2}(a)~and~\ref{fig:figure2}(b).
Otherwise, we obtain closed FSs; see also the Supplemental Materials (SM)~\cite{sm}.
In the real-space representation, the Bogoliubov--de Gennes (BdG) Hamiltonian for the AM segment is given by,
\begin{align}
\begin{split}
&H_{\mathrm{AM}} = \frac{1}{2}\sum_{\boldsymbol{r},\boldsymbol{r}^{\prime} \in \mathrm{AM}}\sum_{\sigma}
\Psi_{\boldsymbol{r}\sigma}^{\dagger}
\left(\begin{array}{cc} \xi_{\boldsymbol{r},\boldsymbol{r}^{\prime}}^{\sigma} & 0\\
0 & -\xi_{\boldsymbol{r},\boldsymbol{r}^{\prime}}^{\bar{\sigma}} \end{array} \right)
\Psi_{\boldsymbol{r}^{\prime}\sigma},\\
&\xi_{\boldsymbol{r},\boldsymbol{r}^{\prime}}^{\sigma} = \varepsilon_{\boldsymbol{r},\boldsymbol{r}^{\prime}}
-\mu_0 \delta_{\boldsymbol{r},\boldsymbol{r}^{\prime}}
+ s_{\sigma} m_{\boldsymbol{r},\boldsymbol{r}^{\prime}},\\
&\varepsilon_{\boldsymbol{r},\boldsymbol{r}^{\prime}}=-\frac{t_0}{2}
\left\{ \delta_{\boldsymbol{r}+\frac{\boldsymbol{x}+\boldsymbol{y}}{2},\boldsymbol{r}^{\prime}}
+ \delta_{\boldsymbol{r}+\frac{\boldsymbol{x}- \boldsymbol{y}}{2},\boldsymbol{r}^{\prime}}
+ ( \boldsymbol{r} \leftrightarrow \boldsymbol{r}^{\prime}) \right\},\\
&m_{\boldsymbol{r},\boldsymbol{r}^{\prime}}=\frac{t_J}{2}
\left\{ \delta_{\boldsymbol{r}+\frac{\boldsymbol{x}+\boldsymbol{y}}{2},\boldsymbol{r}^{\prime}}
- \delta_{\boldsymbol{r}+\frac{\boldsymbol{x}- \boldsymbol{y}}{2},\boldsymbol{r}^{\prime}}
+ ( \boldsymbol{r} \leftrightarrow \boldsymbol{r}^{\prime}) \right\},
\end{split}
\end{align}
where $\Psi_{\boldsymbol{r}\sigma}=(c_{\boldsymbol{r},\sigma}, c_{\boldsymbol{r},\bar{\sigma}}^{\dagger})^{\mathrm{T}}$
with $c_{\boldsymbol{r},\sigma}^{\dagger}$ ($c_{\boldsymbol{r},\sigma}$) representing the creation (annihilation) operator of an electron
at site $\boldsymbol{r}$ with spin $\sigma$,
$\bar{\sigma}$ represents the opposite spin of $\sigma$,
and $\sum_{\boldsymbol{r},\boldsymbol{r}^{\prime}\in \mathrm{AM}}$ denotes the summation over the lattice sites in the AM.
We assume that the lattice sites of the effective AM model are located at $\boldsymbol{r}=j\boldsymbol{x}+m\boldsymbol{y}$
and $\boldsymbol{r}=(j^{\prime}+1/2) \boldsymbol{x}+(m^{\prime}+1/2)\boldsymbol{y}$,
where $\boldsymbol{x}$ ($\boldsymbol{y}$) is the unit vector in the $x$ ($y$) direction.
In the $x$ ($y$) direction, the AM occupies $1 \leq j \leq L$ ($1 \leq m \leq W$) and $1 \leq j^{\prime} \leq L$ ($1 \leq m^{\prime} \leq W-1$).
The lattice configuration of our model is explicitly illustrated in the SM~\cite{sm}.
To describe the two NM leads and the SC, we consider a square lattice, where the lattice site is located at $\boldsymbol{r}=j\boldsymbol{x}+m\boldsymbol{y}$.
The $s$-wave SC occupies the lattice sites with $j>L$ and $1 \leq m \leq W$.
The first (second) NM occupies the lattice sites with $j<1$ and $1 \leq m \leq W_n$ ($W-W_n+1 \leq m \leq W$).
In the $x$ direction, we consider the semi-infinite SC and the semi-infinite NM.
In the $y$ direction, we assume open boundary conditions.
The BdG Hamiltonian for the SC and that for the $\alpha$-th NM are, respectively, given by,
\begin{align}
&H_{\mathrm{SC}} = \frac{1}{2}\sum_{\boldsymbol{r},\boldsymbol{r}^{\prime}\in \mathrm{SC}}\sum_{\sigma}
\Psi_{\boldsymbol{r}\sigma}^{\dagger}
\left(\begin{array}{cc} \xi_{\boldsymbol{r},\boldsymbol{r}^{\prime}} & s_{\sigma}\Delta_{\boldsymbol{r},\boldsymbol{r}^{\prime}}\\
s_{\sigma}\Delta_{\boldsymbol{r},\boldsymbol{r}^{\prime}} & -\xi_{\boldsymbol{r},\boldsymbol{r}^{\prime}} \end{array} \right)
\Psi_{\boldsymbol{r}^{\prime}\sigma},\nonumber\\
&H_{N_{\alpha}} = \frac{1}{2}\sum_{\boldsymbol{r},\boldsymbol{r}^{\prime}\in N_{\alpha}}\sum_{\sigma}
\Psi_{\boldsymbol{r}\sigma}^{\dagger}
\left(\begin{array}{cc} \xi_{\boldsymbol{r},\boldsymbol{r}^{\prime}} & 0\\
0 & -\xi_{\boldsymbol{r},\boldsymbol{r}^{\prime}} \end{array} \right)
\Psi_{\boldsymbol{r}^{\prime}\sigma}, \nonumber\\
&\xi_{\boldsymbol{r},\boldsymbol{r}^{\prime}}=-t\delta_{|\boldsymbol{r}-\boldsymbol{r}^{\prime}|,1}-\mu \delta_{\boldsymbol{r},\boldsymbol{r}^{\prime}},\quad
\Delta_{\boldsymbol{r},\boldsymbol{r}^{\prime}} = \Delta \delta_{\boldsymbol{r},\boldsymbol{r}^{\prime}},
\end{align}
where $t$ is the nearest-neighbor hopping integral, $\mu$ denotes the chemical potential, $\Delta$ is the pair potential,
and $\sum_{\boldsymbol{r},\boldsymbol{r}^{\prime}\in \mathrm{SC}}$ ($\sum_{\boldsymbol{r},\boldsymbol{r}^{\prime}\in N_{\alpha}}$)
represents the summation over the lattice sites in the SC ($\alpha$-th NM) segment.
The interface between the $\alpha$-th NM and the AM and the interface between the AM and the SC are, respectively, described by,
\begin{widetext}
\begin{align}
\begin{split}
&H_{\mathrm{AM}\text{--}N_{\alpha}} = \frac{1}{2}\sum_{m=m_{\alpha}}^{M_{\alpha}}\sum_{\sigma}
\left\{ \Psi_{1,m,\sigma}^{\dagger} \left(\begin{array}{cc} -t^{\mathrm{int}}_{\boldsymbol{r}} & 0\\
0 & t^{\mathrm{int}}_{\boldsymbol{r}} \end{array} \right) \Psi_{0,m,\sigma} + \mathrm{H.c.} \right\},\\
&H_{\mathrm{SC} \text{--}\mathrm{AM}} =  \frac{1}{2}\sum_{m=1}^{W-1}\sum_{\sigma}
\left[ \left\{ \Psi_{L+1,m,\sigma}^{\dagger} \left(\begin{array}{cc} -t^{\mathrm{int}}_{\boldsymbol{r}} & 0\\
0 & t^{\mathrm{int}}_{\boldsymbol{r}} \end{array} \right) \Psi_{L,m+\frac{1}{2},\sigma}
+  \Psi_{L+1,m+1,\sigma}^{\dagger} \left(\begin{array}{cc} -t^{\mathrm{int}}_{\boldsymbol{r}} & 0\\
0 & t^{\mathrm{int}}_{\boldsymbol{r}} \end{array} \right) \Psi_{L,m+\frac{1}{2},\sigma}\right\} + \mathrm{H.c.} \right],
\end{split}
\end{align}
\end{widetext}
where $(m_1,M_1)=(1,W_n)$ and $(m_2,M_2)=(W-W_n+1,W)$.
To effectively describe surface roughness/disorientation,
we include randomness in the hopping integral at the interfaces: $t^{\mathrm{int}}_{\boldsymbol{r}}=t^{\mathrm{int}}+\delta t^{\mathrm{int}}_{\boldsymbol{r}}$,
where $\delta t^{\mathrm{int}}_{\boldsymbol{r}}$ is given randomly in the range of $-\delta t/2 \leq \delta t^{\mathrm{int}}_{\boldsymbol{r}} \leq \delta t/2$.
In the following calculations, we fix the parameters as $t_{0}=t$, $\mu = -t$, $\Delta=0.001t$, $t^{\mathrm{int}}=0.5t$, $W_n=20$, and $W=200$.
As shown in the SM, our main conclusion holds regardless of the details of the parameter choices~\cite{sm}.
Moreover, in the SM, we show that our conclusion remains valid for device geometries with the NMs and the SC on a square lattice tilted by $45^\circ$, as in the AM~\cite{sm};
this shows that our findings are insensitive to the details of the relative lattice orientations at the interfaces.

We calculate the differential conductance $G_{\alpha \beta} = d I_{\alpha}/dV_{\beta}$ in the proposed device.
Here, $I_{\alpha}$ represents the charge current in the $\alpha$-th NM and $V_{\beta}$ denotes the bias voltage applied to the electrode attached to the $\beta$-th NM,
where the electrode attached to the SC is grounded.
Within the Blonder--Tinkham--Klapwijk (BTK) formalism, $G_{\alpha \beta}$ at zero temperature is calculated by~\cite{klapwijk_82,datta_96}
\begin{align}
\begin{split}
&G_{\alpha\beta} = G_{\alpha\beta,\uparrow}+G_{\alpha\beta,\downarrow},\\
&G_{\alpha\beta,\sigma}=\frac{e^2}{h}\mathrm{Tr}\left[ \delta_{\alpha,\beta}\hat{\mathbb{I}}-\hat{R}^e_{\alpha\beta,\sigma}+\hat{R}^h_{\alpha\beta,\sigma}\right],\\
&\hat{R}^{\nu}_{\alpha\beta,\sigma}=\sum_{\sigma^{\prime}=\uparrow,\downarrow}
\hat{s}^{\nu e}_{\alpha\beta,\sigma\sigma^{\prime}} (\hat{s}^{\nu e}_{\alpha\beta,\sigma\sigma^{\prime}})^{\dagger},
\quad (\nu=e,h),
\end{split}
\end{align}
where $\hat{\mathbb{I}}$ is the $N_c \times N_c$ identity matrix, with $N_c$ representing the number of propagating channels per spin in the NM.
The $N_c \times N_c$ matrix of $\hat{s}^{ee}_{\alpha\beta,\sigma\sigma^{\prime}}$ ($\hat{s}^{he}_{\alpha\beta,\sigma\sigma^{\prime}}$) contains
the scattering coefficients from an electron in the $\beta$th NM with spin $\sigma^{\prime}$ to an electron (hole) in the $\alpha$th NM with spin $\sigma$ at energy $E$.
The scattering coefficients are calculated using the recursive Green's function techniques~\cite{fisher_81,ando_91}.
In the BTK formalism, we assume that the currents flowing towards $x=+\infty$ ($x=-\infty$) in the SC (NM) are absorbed into the ideal electrode that are not explicitly described in the Hamiltonian.
The BTK formalism is quantitatively justified for bias voltages well below the superconducting gap.
We also study the zero-frequency noise power defined by $C_{\alpha\beta} = \int^{\infty}_{-\infty} \overline{\delta I_{\alpha}(0) \delta I_{\beta}(t)} dt$,
where $\delta I_{\alpha}(t) = I_{\alpha}(t) - I_{\alpha}$ denotes the deviation of the current at time $t$ from the time averaged current $I_{\alpha}$.
The detailed formulation~\cite{datta_96,beenakker_94} and the numerical results on the zero-frequency noise power are shown in the SM~\cite{sm}.
We note that the transport properties of single crystal RuO$_2$ have already been observed in recent experiments~\cite{liu_22,song_23,veyrat_23,qu_24,jalan_24}.
Furthermore, a very recent experiment has demonstrated the Josephson effect through an AM (although the interface direction is different from that considered in our theory)~\cite{aarts_24}.
On the basis of this rapid progress, we expect that our theory will be verified in near future experiments.

\section{Results}.~%
First, we give a qualitative description of the characteristic scattering processes in the proposed device.
In Fig.~\ref{fig:figure2}(a) [Fig.~\ref{fig:figure2}(b)], we show the FSs of the AM,
which are obtained from the $\Psi_{\uparrow}$-sector ($\Psi_{\downarrow}$-sector) of the Hamiltonian
describing electrons with $\sigma=\uparrow$ ($\downarrow$) and holes with $\sigma=\downarrow$ ($\uparrow$).
The solid (dashed) lines denotes the FSs for the electrons (holes), and the arrows indicate the corresponding group velocity.
The group velocity of the spin-$\sigma$ electrons moving towards the SC segment (i.e., the positive $x$ direction) is approximately given by
$\boldsymbol{v}^{e}_{\sigma,+} \propto (\boldsymbol{x}-s_{\sigma}\boldsymbol{y})$.
The backward waves of spin-$\sigma$ electrons and spin-$\bar{\sigma}$ holes have the group velocities,
$\boldsymbol{v}^{e}_{\sigma,-} \propto (-\boldsymbol{x}+s_{\sigma}\boldsymbol{y})$
and $\boldsymbol{v}^{h}_{\bar{\sigma},-} \propto (-\boldsymbol{x}-s_{\sigma}\boldsymbol{y})$, respectively.
Here, we consider that an electron is injected from the first NM, as shown in Figs.~\ref{fig:figure2}(c) and \ref{fig:figure2}(d).
As shown in Fig.~\ref{fig:figure2}(c), the spin-$\uparrow$ electron with $\boldsymbol{v}^{e}_{\uparrow,+} \propto (\boldsymbol{x}-\boldsymbol{y})$ in the AM
is scattered at the bottom boundary of the AM.
Since $\boldsymbol{v}^{e}_{\uparrow,+} \approx - \boldsymbol{v}^{e}_{\uparrow,-}$, the scattered electron traces back the original trajectory of the incident electron.
This retroreflective nature of \emph{normal} reflections in the AM leads to $R^{e}_{21,\uparrow}=0$,
where $R^{\nu}_{\alpha\beta,\sigma} = \mathrm{Tr}[\hat{R}^{\nu}_{\alpha\beta,\sigma}]$.
In contrast, as shown in Fig.~\ref{fig:figure2}(d),
the spin-$\downarrow$ electron having $\boldsymbol{v}^{e}_{\downarrow,+} \propto (\boldsymbol{x}+\boldsymbol{y})$ reaches the AM--SC interface.
Remarkably, since $\boldsymbol{v}^{e}_{\downarrow,+}$ and $\boldsymbol{v}^{h}_{\uparrow,-}$ differ only in the sign of the $x$ component,
the incident spin-$\downarrow$ electron undergoes the SAR to the spin-$\uparrow$ hole, which leads to $R^{h}_{21,\uparrow}\neq 0$.
In the absence of spin-flip scatterings, we also obtain $R^{e}_{21,\uparrow}=R^{h}_{21,\downarrow}=0$.
A definitive observable signature of the SAR in the proposed device is the positive nonlocal conductance, $G_{21}>0$,
which implies $\sum_{\sigma}R^{h}_{21,\sigma}>\sum_{\sigma}R^{e}_{21,\sigma}$.
Moreover, since only the SAR in Fig.~\ref{fig:figure2}(d) contributes to the charge currents, we expect $G_{21,\uparrow}=G_{21}=G_{11,\downarrow}=G_{11}=(e^2/h) R^{h}_{21,\uparrow}$,
indicating that the current in the first (second) NM is fully polarized to $\downarrow$-spin ($\uparrow$-spin).
When we inject the electron from the second NM, as illustrated in Figs.~\ref{fig:figure2}(e) and \ref{fig:figure2}(f),
we expect the occurrence of SAR ($R^{h}_{12,\downarrow}\neq 0$) and the absence of other inter-lead scatterings ($R^{e}_{12,\downarrow}=R^{h}_{12,\uparrow}=R^{h}_{12,\uparrow}=0$),
which leads to $G_{12,\downarrow}=G_{12}=G_{22,\uparrow}=G_{22}$.
Remarkably, the reversal scattering process of SARs in Figs.~\ref{fig:figure2}(d) and \ref{fig:figure2}(e) correspond to the Cooper pair splitting as illustrated in Fig.~\ref{fig:figure2}(g):
a Cooper pair injected from the SC splits into a spin-$\downarrow$ electron flowing into the first NM and the spin-$\uparrow$ electron flowing into the second NM.
Therefore, the proposed device inherently functions as a Cooper pair splitter equipped with the spin filters.

\begin{figure}[b]
\begin{center}
\includegraphics[width=0.5\textwidth]{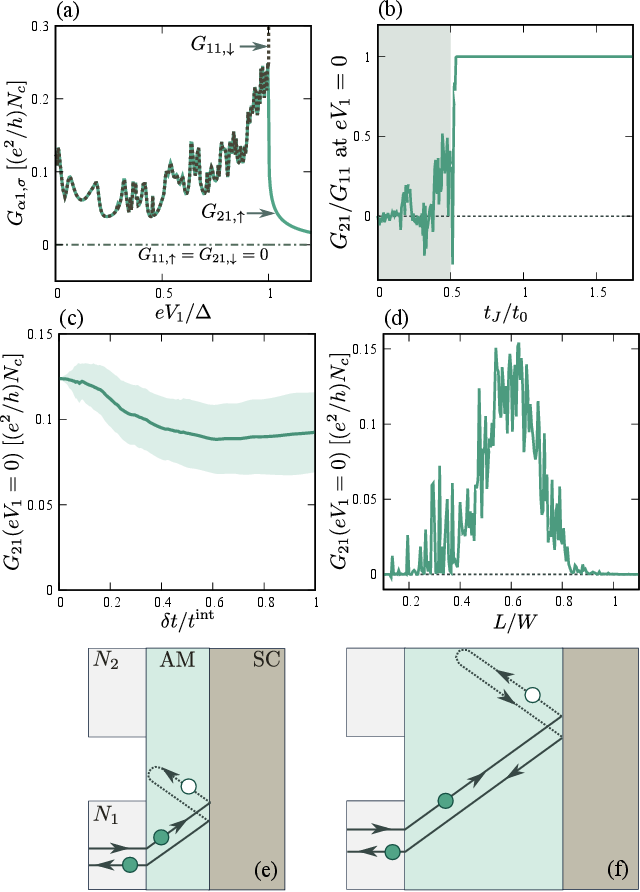}
\caption{(a) Differential conductance of $G_{\alpha1,\sigma}$ as a function of the bias voltage.
For $eV<\Delta$, the relation $G_{21,\uparrow}=G_{11,\downarrow}=G_{21}=G_{11}$ holds.
(b) $G_{21}/G_{11}$ at zero-bias voltage as a function of the exchange potential $t_J$.
In the shaded region for $t_J \leq 0.5t_0$,  the AM does not exhibit the spin-split open FSs.
(c) $G_{21}$ at zero bias voltage as a function of the strength of the roughness at the interface (i.e., $\delta t$).
The solid line (shaded area) represents the averaged value (standard deviation) for 100 samples.
(d) $G_{21}$ at zero-bias voltage as a function of $L/W$.
(e), (f) Failure of SARs, which explains $G_{21}\approx 0$ for the smaller/larger ratio of $L/W$.}
\label{fig:figure3}
\end{center}
\end{figure}
Then, we numerically verify the above expectation.
In Fig.~\ref{fig:figure3}(a), we show the differential conductance of $G_{\alpha1,\sigma}$ as a function of $eV_1$,
where the parameters are chosen as $(t_J,\mu_0,L) = (0.75t,0.5t,95)$.
For simplicity, we use a uniform hopping integral at the interfaces, i.e., $\delta t=0$.
The result is normalized by $(e^2/h) N_c$, where $N_c=14$ with the present parameters.
In perfect agreement with our expectation, we obtain $G_{11,\downarrow}=G_{21,\uparrow}>0$ for $eV_1<\Delta$, while other components are zero.
Namely, we numerically confirm the relation of $G_{21,\uparrow}=G_{21}=G_{11,\downarrow}=G_{11}=(e^2/h) R^{h}_{21,\uparrow}$.
The oscillation of $G_{\alpha1,\sigma}$ for $eV_1<\Delta$ suggests the presence of quantum interference effects due to the confinement in the AM segment~\cite{stone_90}.
In Fig.~\ref{fig:figure3}(b), we show $G_{21}/G_{11}$ at zero-bias voltage as a function of the exchange potential $t_J$, where $(\mu_0,\delta t,L) = (t,0,95)$.
In the shaded region for $t_J \leq 0.5t_0$,  the AM does not exhibit the spin-split open FSs.
We find that the nonlocal conductance stably becomes positive and satisfies $G_{21}=G_{11}$ when the AM exhibits the spin-split open FSs.
In Fig.~\ref{fig:figure3}(c), we show $G_{21}$ as a function of the strength of the randomness at the interface (i.e., $\delta t$), where $(t_J,\mu_0,L) = (0.75t,0.5t,95)$.
The solid line (shaded area) represents the averaged value (standard deviation) for 100 samples.
We find that the positive nonlocal conductance remains even in the presence of the surface randomness.
Consequently, we conclude that the SAR, experimentally detected through the positive nonlocal conductance, occurs over a broad parameter space,
as long as the AM exhibits the spin-split open FSs.
Lastly, in Fig.~\ref{fig:figure3}(c), we show $G_{21}$ at zer- bias voltage as a function of the length of the AM (i.e., $L$), where we choose $(t_J,\mu_0,\delta t) = (0.75t,0.5t,0)$.
The nonlocal conductance $G_{21}$ takes substantial positive values in the wide range of the parameter around $L/W=0.6$.
The absence of $G_{21}$ for shorter and longer $L$ is intuitively understood by the scattering processes illustrated in Fig.~\ref{fig:figure3}(d) and Fig.~\ref{fig:figure3}(e).
Namely, in a system with an inappropriately small/large ratio of $L/W$,
a specularly Andreev reflected hole undergoes the retronormal reflection at the surfaces of the AM before reaching the second NM and eventually returns to the first NM as an electron state.
Therefore, the system configuration, especially the ratio of $L/W$, is an important factor to observe the SAR in the AM--SC hybrid device.
In the SM, we also demonstrate the positive noise cross-correlation, $C_{12}=C_{21}>0$, for $eV_1=eV_2<\Delta$,
which is another smoking-gun signature of the SAR and the Cooper pair splitting in the proposed device~\cite{martin_99}.

\section{Discussion}.~%
In realistic AM materials, local noncentrosymmetry can induce asymmetric spin-orbit coupling
that perturbatively hybridizes spin-up and spin-down states and disturbs the open FSs~\cite{smejkal_20}.
In the SM, we effectively examine the effects of such weak asymmetric spin-orbit coupling within our single-band model
and describe the emergence of the positive nonlocal conductance under such perturbations~\cite{sm}.
As described in Fig.~\ref{fig:figure2}, the occurrence of SARs is entirely due to the unique configuration of Fermi surfaces in altermagnets (i.e., spin-splitting open Fermi surfaces),
where such Fermi surfaces are also described through microscopic calculations~\cite{elmers_24,qian_24}.
Therefore, we can expect that SARs occur insensitive to the specific details of the model.
Nevertheless, further quantitative calculations based on more realistic microscopic models will be a crucial future task.

In summary, we demonstrate that the characteristic spin-split band of the AM leads to the occurrence of spin-polarized SARs at the AM--SC interfaces.
By utilizing this scattering process, we propose a multiterminal device that integrates the functionalities of both a Cooper pair splitter and a spin beam splitter,
thereby paving a promising way for the creation of energy-entangled electron pairs.
The positive nonlocal conductance and positive noise-correlation serve as smoking-gun signatures of SARs in the proposed device.

\begin{acknowledgments}
\textit{Acknowledgements}.~
We thank Y. Maeno for the insightful discussions. 
S.I. is supported by the Grant-in-Aid for JSPS Fellows (JSPS KAKENHI Grant No. JP22KJ1507) and for Early-Career Scientists (JSPS KAKENHI Grant No. JP24K17010).
\end{acknowledgments}

\section*{Data availability}
The data that support the findings of this article are openly available \cite{data}.

\clearpage

\onecolumngrid
\begin{center}
  \textbf{\large Supplemental Material for \\ ``Spin-polarized Specular Andreev Reflections in Altermagnets''}\\ \vspace{0.3cm}
Yutaro Nagae$^{1}$, Andreas P. Schnyder$^{2}$, and Satoshi Ikegaya$^{1,3}$\\ \vspace{0.1cm}
{\itshape $^{1}$Department of Applied Physics, Nagoya University, Nagoya 464-8603, Japan\\
$^{2}$Max-Planck-Institut f\"ur Festk\"orperforschung, Heisenbergstrasse 1, D-70569 Stuttgart, Germany\\
$^{3}$Institute for Advanced Research, Nagoya University, Nagoya 464-8601, Japan}
\date{\today}
\end{center}

\section{Fermi surfaces of altermagnet}
\begin{figure}[h]
\begin{center}
\includegraphics[width=0.75\textwidth]{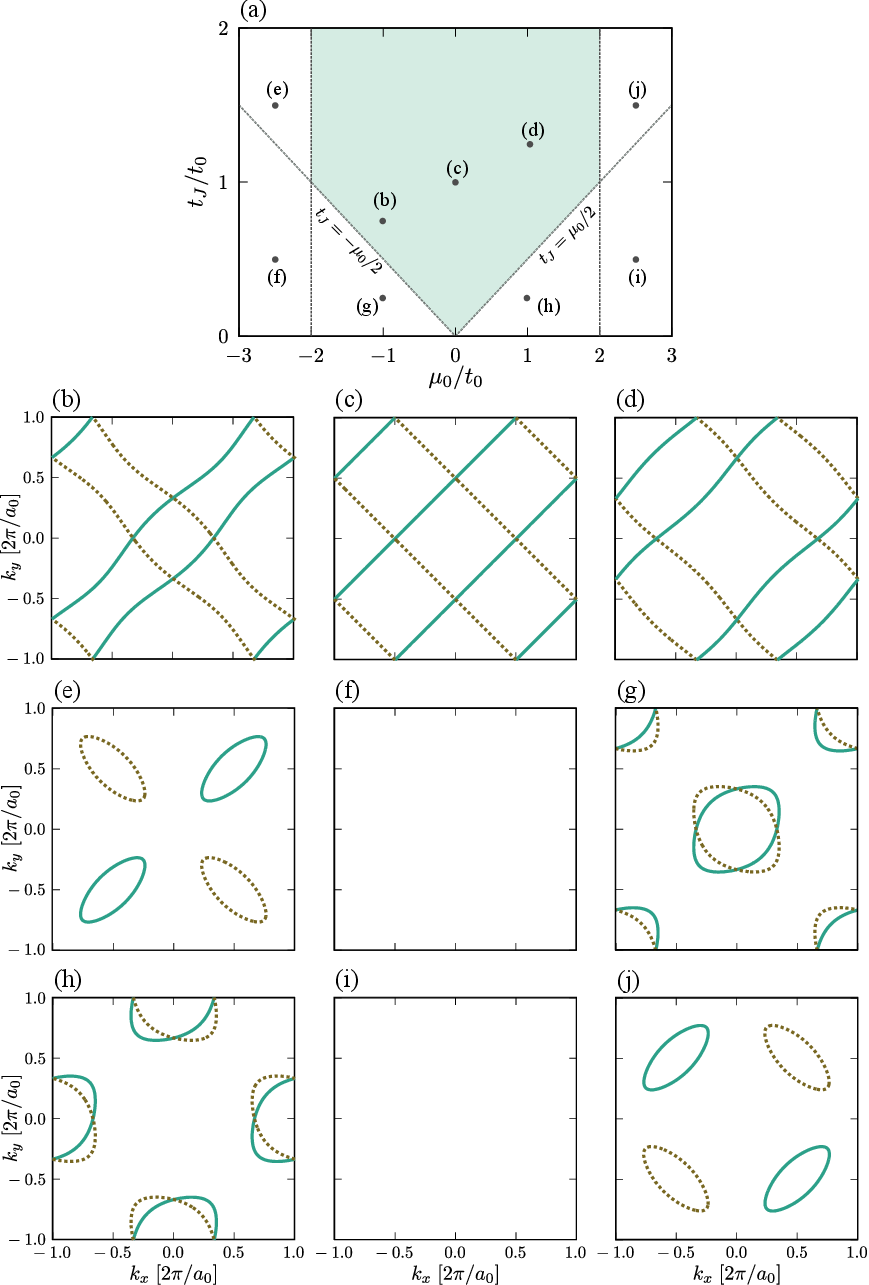}
\caption{(a) Phase diagram for the configurations of the FSs in the AM.
In the shaded region, where $|\mu_0|<2t_J$ with $t_J \leq t_0$ or $|\mu_0|<2t_0$ with $t_J \geq t_0$ are satisfied, the AM exhibits the spin-split open FSs.
(b)-(j) FSs at $(t_J/t_0,\mu_0/t_0)=(0.75,-1.0),\;(1.0,0.0),\;(1.25,1.0),\;(1.5,-2.5),\;(0.5,-2.5),\;(0.25,-1.0),\;(0.25,1.0),\;(0.5,2.5),\;(1.5,2.5)$, respectively.
The parameters are also indicated in (a).
We find the spin-split open FSs in Figs.~(b)-(d).
In Fig.~(f) and Fig.~(i), FSs are absent.}
\label{fig:figure1_sp}
\end{center}
\end{figure}

In this section, we discuss the Fermi surfaces (FSs) of an altermagnet (AM), described by the effective single-band Hamiltonian in Eq.~(1) of the main text~\cite{smejkal_22(1),linder_23(2),linder_23(3)}:
\begin{align}
\begin{split}
&H_{\mathrm{AM}}=\sum_{\boldsymbol{k},\sigma} \left( \varepsilon_{\boldsymbol{k}} - \mu_0
+ s_{\sigma} m_{\boldsymbol{k}} \right) c^{\dagger}_{\boldsymbol{k}\sigma}c_{\boldsymbol{k}\sigma},\\
&\varepsilon_{\boldsymbol{k}}=-2t_0 \cos \frac{k_x}{2} \cos\frac{k_y}{2}, \\
&m_{\boldsymbol{k}} = -2t_J \sin\frac{k_x}{2} \sin\frac{k_y}{2},
\end{split}
\end{align}
where $c_{\boldsymbol{k}\sigma}^{\dagger}$ ($c_{\boldsymbol{k}\sigma}$) is the creation (annihilation) operator of an electron
with momentum $\boldsymbol{k}$ and spin $\sigma$ ($\uparrow$ or $\downarrow$),
$t_0>0$ denotes the nearest-neighbor hopping integral on a square-lattice oriented at a $45$ degree angle,
$\mu_0$ is the chemical potential, $t_J>0$ is the strength of the $d_{xy}$-wave exchange potential, and $s_{\sigma} = +1 (-1)$ for $\sigma=\uparrow$ ($\downarrow$).
To examine the FSs, we focus on the energy spectrum of the spin-$\uparrow$ states:
\begin{align}
\xi_{\boldsymbol{k},\uparrow} = \varepsilon_{\boldsymbol{k}} - \mu_0 + m_{\boldsymbol{k}},
\end{align}
where the FSs are located at momenta satisfying $\xi_{\boldsymbol{k},\uparrow}=0$.
For $t_0=t_J$, we obtain
\begin{align}
\xi_{\boldsymbol{k},\uparrow} =-2t_0 \cos\left( \frac{k_x-k_y}{2} \right) - \mu_0
\end{align}
and clearly find that the FSs has an open shape along the $(k_x+k_y)$-direction.
This open FSs remain as long as the following three conditions are satisfied:
(i) the presence of momenta satisfying $\xi_{k_x,k_y=2\pi,\uparrow}=0$ (i.e., the FSs intersects the Brillouin zone boundary),
(ii) the presence of momenta satisfying $\xi_{k,-k,\uparrow}=0$ (i.e., the FSs intersect the line along the $(k_x-k_y)$-direction),
and (iii) the absence of momenta satisfying $\xi_{k,k,\uparrow}=0$ (i.e., the FSs do not exist at momenta satisfying $k_x=k_y$).
These conditions lead to:
\begin{align}
\begin{split}
&\text{(i): }|\mu_0| < 2t_0,\\
&\text{(ii): } 0<\frac{2t_0+\mu_0}{2t_0+2t_J}<1,\\
&\text{(iii): } \frac{2t_0+\mu_0}{2t_0-2t_J}<0 \quad \text{or} \quad \frac{2t_0+\mu_0}{2t_0-2t_J}>1.
\label{eq:fs_condition}
\end{split}
\end{align}
By summarizing Eq.~(\ref{eq:fs_condition}), we obtain the phase diagram in Fig.~\ref{fig:figure1_sp}(a).
The open FSs appear in the shaded region, where the conditions of $|\mu_0|<2t_J$ with $t_J \leq t_0$ or $|\mu_0|<2t_0$ with $t_J \geq t_0$ are satisfied.
We also show the typical configurations of the FSs in Figs.~\ref{fig:figure1_sp}(b)-\ref{fig:figure1_sp}(h).
\clearpage

\section{Specular Andreev reflection under various conditions}
In this section, we discuss the appearance of the positive nonlocal conductance under various conditions, which serves as a definitive signature of the specular Andreev reflection in the AM.
\subsection{Insensitivity to the parameters in the Hamiltonian}
\begin{figure}[h]
\begin{center}
\includegraphics[width=0.9\textwidth]{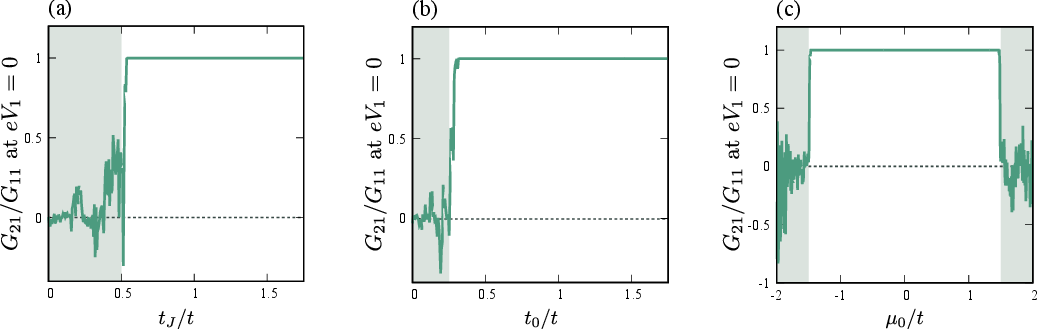}
\caption{(a) $G_{21}/G_{11}$ at zero bias voltage as a function of $t_J$ with $(t_0,\mu_0)=(t,t)$.
(b) $G_{21}/G_{11}$ at zero bias voltage as a function of $t_0$ with $(t_J,\mu_0)=(0.75t,0.5t)$.
(c)$G_{21}/G_{11}$ at zero bias voltage as a function of $\mu_0$ with $(t_0,t_J)=(t,0.75t)$.
In the shaded regions, the AM does not exhibit the spin-split open FSs.}
\label{fig:figure2_sp}
\end{center}
\end{figure}

In Fig.~\ref{fig:figure2_sp}, we show $G_{21}/G_{11}$ at zero-bias voltage as a functions of the parameters in $H_{\mathrm{AM}}$,
where we use the tight-binding Hamiltonian shown in the main text and fix the parameters as $\mu = -t$, $\Delta=0.001t$, $t^{\mathrm{int}}=0.5t$, $W_n=20$, $W=200$, and $L=95$.
For simplicity, in this Supplemental Material, we assume the uniform hopping integral at the interfaces, i.e., $\delta t=0$.
The result is normalized by $(e^2/h) N_c$, where $N_c=14$ with the present parameters.
In Fig.~\ref{fig:figure2_sp}(a), we show $G_{21}/G_{11}$ as a function of $t_J$ with $(t_0,\mu_0)=(t,t)$, which is equivalent to that in Fig.~3(b) of the main text.
In Fig.~\ref{fig:figure2_sp}(b), we show $G_{21}/G_{11}$ as a function of $t_0$ with $(t_J,\mu_0)=(0.75t,0.5t)$.
In Fig.~\ref{fig:figure2_sp}(c), we show $G_{21}/G_{11}$ as a function of $\mu_0$ with $(t_0,t_J)=(t,0.75t)$.
The shaded regions in Fig.~\ref{fig:figure2_sp}, correspond to the parameter region where the AM does not exhibit the open FSs.
As a result, we confirm the realization of the positive nonlocal conductance over a broad parameter space where the AM exhibits the spin-split open FSs.
\clearpage

\subsection{Insensitivity to the relative lattice orientations at the interfaces}
\begin{figure}[h]
\begin{center}
\includegraphics[width=0.85\textwidth]{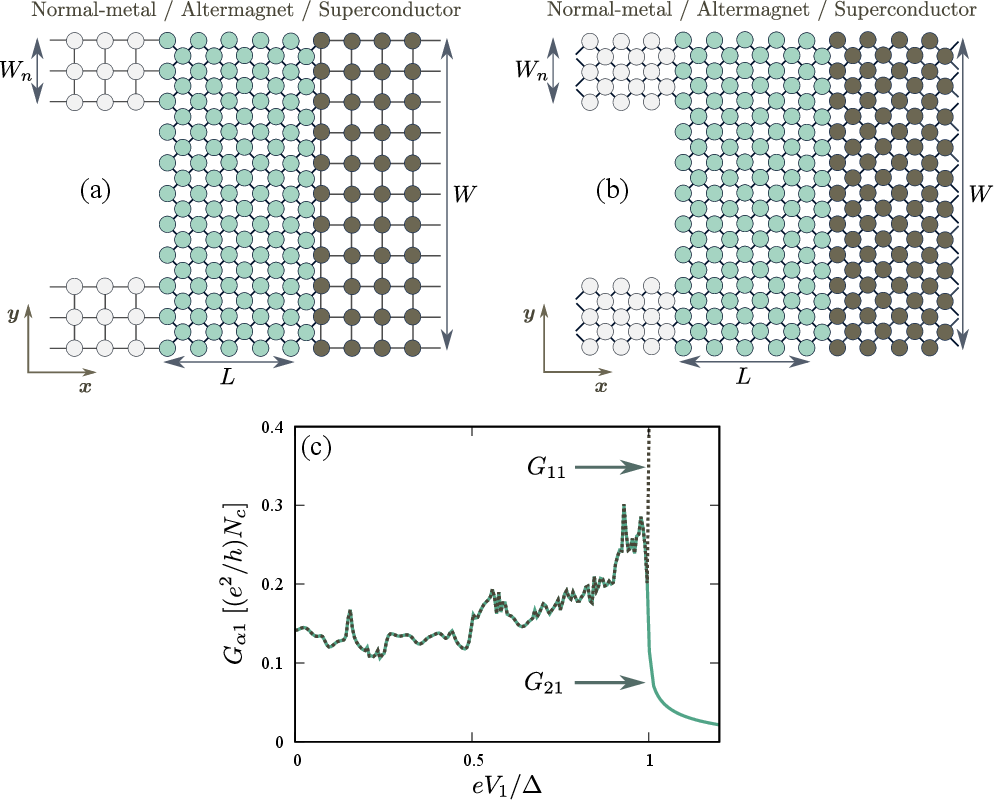}
\caption{(a) Device geometry used in the main text.
(b) Device geometry, where all SC, NM, and AM segments are described on the square lattice orientated at a 45 degree angle.
(c) Differential conductance of $G_{\alpha1}$ in the device (b) as function of the bias voltage $eV_1$.}
\label{fig:figure3_sp}
\end{center}
\end{figure}

In the main text, we use the square lattice in the superconductor (SC) and the normal metal (NM) segments,
while the AM is described on the square lattice orientated at a 45 degree angle, as shown in Fig.~\ref{fig:figure3_sp}(a).
In this section, we discus the nonlocal conductance in the device where all SC, NM, and AM segments are described on the square lattice orientated at a 45 degree angle,
as shown in Fig.~\ref{fig:figure3_sp}(b).
To describe the present junction, the Hamiltonian for the SC is replaced with
\begin{align}
\begin{split}
&H_{\mathrm{SC}} = \frac{1}{2}\sum_{\boldsymbol{r},\boldsymbol{r}^{\prime}\in \mathrm{SC}}\sum_{\sigma}
\Psi_{\boldsymbol{r}\sigma}^{\dagger}
\left(\begin{array}{cc} \xi^{\prime}_{\boldsymbol{r},\boldsymbol{r}^{\prime}} & s_{\sigma}\Delta_{\boldsymbol{r},\boldsymbol{r}^{\prime}}\\
s_{\sigma}\Delta_{\boldsymbol{r},\boldsymbol{r}^{\prime}} & -\xi^{\prime}_{\boldsymbol{r},\boldsymbol{r}^{\prime}} \end{array} \right)
\Psi_{\boldsymbol{r}^{\prime}\sigma},\nonumber\\
&\xi^{\prime}_{\boldsymbol{r},\boldsymbol{r}^{\prime}}=-\frac{t}{2}
\left\{ \delta_{\boldsymbol{r}+\frac{\boldsymbol{x}+\boldsymbol{y}}{2},\boldsymbol{r}^{\prime}}
+ \delta_{\boldsymbol{r}+\frac{\boldsymbol{x}- \boldsymbol{y}}{2},\boldsymbol{r}^{\prime}}
+ ( \boldsymbol{r} \leftrightarrow \boldsymbol{r}^{\prime}) \right\}-\mu \delta_{\boldsymbol{r},\boldsymbol{r}^{\prime}},\\
&\Delta_{\boldsymbol{r},\boldsymbol{r}^{\prime}} = \Delta \delta_{\boldsymbol{r},\boldsymbol{r}^{\prime}},
\end{split}
\end{align}
where  $t$ is the nearest-neighbor hopping integral, $\mu$ denotes the chemical potential, $\Delta$ is the pair potential,
and $\sum_{\boldsymbol{r},\boldsymbol{r}^{\prime}\in \mathrm{SC}}$ represents the summation over the lattice sites in the SC segment.
We assume that the lattice sites are located at $\boldsymbol{r}=j\boldsymbol{x}+m\boldsymbol{y}$
and $\boldsymbol{r}=(j^{\prime}+1/2) \boldsymbol{x}+(m^{\prime}+1/2)\boldsymbol{y}$,
where $\boldsymbol{x}$ ($\boldsymbol{y}$) is the unit vector in the $x$ ($y$) direction.
In the $x$ ($y$) direction, the SC occupies $L+1 \leq j $ ($1 \leq m \leq W$) and $L+1 \leq j^{\prime}$ ($1 \leq m^{\prime} \leq W-1$).
The $\alpha$-th NM is described by
\begin{align}
&H_{N_{\alpha}} = \frac{1}{2}\sum_{\boldsymbol{r},\boldsymbol{r}^{\prime}\in N_{\alpha}}\sum_{\sigma}
\Psi_{\boldsymbol{r}\sigma}^{\dagger}
\left(\begin{array}{cc} \xi^{\prime}_{\boldsymbol{r},\boldsymbol{r}^{\prime}} & 0\\
0 & -\xi^{\prime}_{\boldsymbol{r},\boldsymbol{r}^{\prime}} \end{array} \right)
\Psi_{\boldsymbol{r}^{\prime}\sigma},
\end{align}
where $\sum_{\boldsymbol{r},\boldsymbol{r}^{\prime}\in N_{\alpha}}$ represents the summation over the lattice sites belonging to the $\alpha$-th NM.
In the $x$-direction, the NMs occupy $j\leq0$ and $j^{\prime}\leq0$.
In the $y$-direction, the first (second) NM occupies  $1 \leq m \leq W_n$ ($W-W_n+1 \leq m \leq W$) and $1 \leq m^{\prime} \leq W_n-1$ ($W-W_n+1 \leq m^{\prime} \leq W-1$).
The interface between the $\alpha$-th NM and the AM and the interface between the AM and the SC are, respectively, described by
\begin{align}
\begin{split}
&H_{\mathrm{AM}\text{--}N_{\alpha}} = \frac{1}{2}\sum_{m=m_{\alpha}}^{M_{\alpha}}\sum_{\sigma}
\left\{ \Psi_{1,m,\sigma}^{\dagger} \left(\begin{array}{cc} -t^{\mathrm{int}} & 0\\
0 & t^{\mathrm{int}} \end{array} \right) \Psi_{0,m+\frac{1}{2},\sigma} + \mathrm{H.c.} \right\},\\
&H_{\mathrm{SC} \text{--}\mathrm{AM}} =  \frac{1}{2}\sum_{m=1}^{W-1}\sum_{\sigma}
\left[ \left\{ \Psi_{L+1,m,\sigma}^{\dagger} \left(\begin{array}{cc} -t^{\mathrm{int}} & 0\\
0 & t^{\mathrm{int}} \end{array} \right) \Psi_{L,m+\frac{1}{2},\sigma}
+  \Psi_{L+1,m+1,\sigma}^{\dagger} \left(\begin{array}{cc} -t^{\mathrm{int}} & 0\\
0 & t^{\mathrm{int}} \end{array} \right) \Psi_{L,m+\frac{1}{2},\sigma}\right\} + \mathrm{H.c.} \right],
\end{split}
\end{align}
where $(m_1,M_1)=(1,W_n-1)$ and $(m_2,M_2)=(W-W_n+1,W-1)$.
For simplicity, we use the uniform hopping integral at the interfaces.
In Fig.~\ref{fig:figure3_sp}(c), we show the nonlocal conductance $G_{21}$ as a function of the bias voltage.
We choose parameters as $\mu = -t$, $\Delta=0.001t$, $t_0=t$, $t_J=0.75t$, $\mu_0=0.5t$, $t^{\mathrm{int}}=0.5t$, $W_n=20$, $W=200$, and $L=95$.
The result is normalized by $(e^2/h) N_c$, where $N_c=16$ with the present parameters.
We find the positive nonlocal conductance $G_{21}>0$ for $eV_1<\Delta$ even for device geometries with the NMs and the SC on a square lattice tilted by $45$ degrees.
In addition, we obtain $G_{11}=G_{21}$ for $eV_1<\Delta$, which implies only the specular Andreev reflection contributes to the charge currents.
Note that, in the main text, we also demonstrate the robust realization of the positive nonlocal conductance in the presence of surface roughness.
These results suggest that our findings are insensitive to the details of the relative lattice orientations at the interfaces.
\clearpage

\subsection{Effects of asymmetric spin-orbit couplings}
\begin{figure}[h]
\begin{center}
\includegraphics[width=0.4\textwidth]{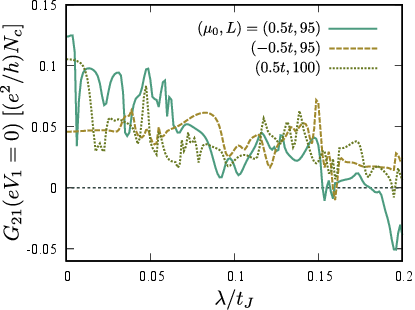}
\caption{Nonlocal conductance $G_{21}$ at zero-bias voltage as a function of the strength of the spin-orbit coupling potential (i.e., $\lambda$).
The results for $(\mu_0,L) = (0.5t_0,95)$,~$(-0.5t_0,95)$,~and~$(0.5t_0,100)$ are plotted.}
\label{fig:figure4_sp}
\end{center}
\end{figure}

In realistic AM materials, local noncentrosymmetry can induce asymmetric spin-orbit coupling that perturbatively hybridizes spin-up and spin-down states
and disturbs the spin-split open Fermi surfaces~\cite{smejkal_20}.
In this section, we examine the effects of such weak asymmetric spin-orbit coupling against the positive nonlocal conductance.
In order to effectively describe asymmetric spin-orbit couplings within our single-band minimal model,
we alternatively employ a typical asymmetric spin-orbit coupling in two-dimensional single-band systems, namely the Rashba spin-orbit coupling:
\begin{align} 
\begin{split}
&H_R = \sum_{\boldsymbol{r}\in \mathrm{AM}}\sum_{\sigma, \sigma^{\prime}} ( \Lambda_{\boldsymbol{r},\sigma, \sigma^{\prime}}+ \mathrm{H.c.}), \\
&\Lambda_{\boldsymbol{r},\sigma, \sigma^{\prime}}
= -i\frac{\lambda}{2} \sum_{s=\pm} \left\{ \left( \boldsymbol{z} \times \sqrt{2}\boldsymbol{d}_s \right) \cdot \boldsymbol{\sigma}\right\}_{\sigma \sigma^{\prime}}
c^{\dagger}_{\boldsymbol{r}+\boldsymbol{d}_s,\sigma} c_{\boldsymbol{r},\sigma^{\prime}},
\end{split}
\end{align}
where $\boldsymbol{z}$ represents the unit vector in the $z$-direction,
$\boldsymbol{d}_{\pm} = (\boldsymbol{x}\pm\boldsymbol{y})/2$, and $\boldsymbol{\sigma}=(\sigma_x,\sigma_y,\sigma_z)$ denotes the Pauli matrices in spin space.
We note that this approach has also been used in previous studies~[\onlinecite{smejkal_22(0)}]. 
In Fig.~\ref{fig:figure4_sp}, we show $G_{21}$  at zero-bias voltage as a function of the strength of the spin-orbit coupling $\lambda$,
where the parameters are chosen as $\mu = -t$, $\Delta=0.001t$, $t_0=t$, $t_J=0.75t$, $t^{\mathrm{int}}=0.5t$, $\delta t=0$, $W_n=20$, $W=200$.
The number of propagating channels in the NM is given by $N_c=14$.
We show the results for $(\mu_0,L) = (0.5t_0,95)$,~$(-0.5t_0,95)$,~and~$(0.5t_0,100)$.
We find that $G_{21}$ shows non-monotonic behavior as a function of $\lambda$, which may arise from complicated quantum interference effects in the AM region~\cite{stone_90}.
For large $\lambda$, the nonlocal conductance can be negative, which  implies $\sum_{\sigma}R^{h}_{21,\sigma}<\sum_{\sigma}R^{e}_{21,\sigma}$,
where the \emph{specular} normal reflections are induced by spin-flip scattering in the presence of the spin-orbit coupling.
Nevertheless, we find that the nonlocal conductance remains positive for $\lambda$ values up to about $0.15 t_J$.
These results suggest that the positive nonlocal conductance remains even in the presence of the \textit{perturbative} spin-orbit couplings, i.e., $\lambda \ll t_J$.
We also note that the study of transport properties based on more realistic microscopic models remains an important future task.
\clearpage

\section{Positive noise cross-correlation}
\begin{figure}[h]
\begin{center}
\includegraphics[width=0.4\textwidth]{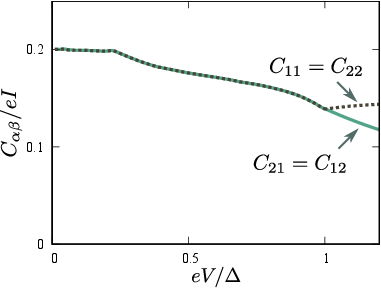}
\caption{Zero-frequency noise power as a function of bias voltage, where $eV_1=eV_2=eV$.}
\label{fig:figure5_sp}
\end{center}
\end{figure}

In this section, we discuss the zero-frequency noise power of the present junction,
defined by $C_{\alpha\beta} = \int^{\infty}_{-\infty} \overline{\delta I_{\alpha}(0) \delta I_{\beta}(t)} dt$,
where $\delta I_{\alpha}(t) = I_{\alpha}(t) - I_{\alpha}$ denotes the deviation of the current at time $t$ from $I_{\alpha}$ representing the time averaged current in the $\alpha$th NM.
Within the Blonder--Tinkham--Klapwijk formalism, the zero-frequency noise power at zero-temperature is calculated by~\cite{klapwijk_82,beenakker_94,datta_96}
\begin{align}
\begin{split}
&C_{\alpha\beta} = \frac{e^2}{h}\int^{eV}_{0} P_{\alpha\beta} dE,\\
&P_{\alpha\beta} = \mathrm{Tr}\left[ \delta_{\alpha\beta}\sum_{\nu=e,h}\hat{Q}^{\nu\nu}_{\alpha\alpha}
- \sum_{\nu,\nu^{\prime}}\sigma_{\nu}\sigma_{\nu^{\prime}}\hat{Q}^{\nu\nu^{\prime}}_{\alpha\beta}\hat{Q}^{\nu^{\prime}\nu}_{\beta\alpha}\right],\\
&\hat{Q}^{\nu\nu^{\prime}}_{\alpha\beta}=\sum_{\gamma=1,2} \hat{s}^{\nu e}_{\alpha\gamma} (\hat{s}^{\nu^{\prime} e}_{\beta\gamma})^{\dagger},\\
&\hat{s}^{\nu e}_{\alpha\beta}=\left(\begin{array}{cc} \hat{s}^{\nu e}_{\alpha\beta,\uparrow\uparrow}&\hat{s}^{\nu e}_{\alpha\beta,\uparrow\downarrow}\\
\hat{s}^{\nu e}_{\alpha\beta,\downarrow\uparrow}&\hat{s}^{\nu e}_{\alpha\beta,\downarrow\downarrow} \end{array}\right),
\end{split}
\end{align}
where $\sigma_{\nu}=1$ ($-1$) for $\nu=e$ ($h$).
The $N_c \times N_c$ matrix of $\hat{s}^{ee}_{\alpha\beta,\sigma\sigma^{\prime}}$ ($\hat{s}^{he}_{\alpha\beta,\sigma\sigma^{\prime}}$) contains
the scattering coefficients from an electron in the $\beta$-th NM with spin $\sigma^{\prime}$ to an electron (hole) in the $\alpha$-th NM with spin $\sigma$ at energy $E$.
In Fig.~\ref{fig:figure5_sp}, we show the zero-frequency noise power, $C_{\alpha\beta}$, as a function of the bias voltage,
where we apply the same bias voltage to both leads, i.e., $eV_1=eV_2=eV$.
We choose $\mu = -t$, $\Delta=0.001t$, $t_0=t$, $t_J=0.75t$, $\mu_0=0.5t$, $t^{\mathrm{int}}=0.5t$, $\delta t=0$, $W_n=20$, $W=200$, and $L=95$.
The results are normalized by $eI=e(I_1+I_2)$.
We clearly find a positive cross-correlation of $C_{12}=C_{21}>0$, which is an important signature of the Cooper pair splitting~\cite{martin_99}.
In addition, the relation of $2C_{12} = C_{11}+C_{22}$ holds for $eV<\Delta$,
where the cross-correlation in any stochastic process is bounded by  the auto-correlation as $2|C_{12}| \leq C_{11}+C_{22}$.
Therefore, we find that the specular Andreev reflection in the AM induces remarkably strong cross-correlations between the currents following in the different NM leads.

\end{document}